# L-TGVN: Leveraging Longitudinal Priors for Personalized Rapid MRI

**Arda Atalık**[1,2], **Sumit Chopra**[2,3], **Daniel K. Sodickson**[2,4]
[1] NYU Center for Data Science, NY, USA
[2] Center for Advanced Imaging Innovation and Research (CAI²R), Department of Radiology, NYU Grossman School of Medicine, NY, USA
[3] Courant Institute of Mathematical Sciences, NY, USA
[4] Function Health, TX, USA
{Arda.Atalik, Sumit.Chopra, Daniel.Sodickson}@nyulangone.org

**Abstract**

MRI provides excellent soft-tissue contrast without ionizing radiation, but long acquisition times increase patient discomfort while also raising exam costs and limiting scanner throughput. A common approach to reduce scan time is to acquire fewer measurements, which yields an ill-posed linear inverse problem; recovering diagnostic-quality images therefore requires incorporating prior knowledge beyond the measured data. In follow-up exams, the most recent prior scan of a patient can provide a highly informative subject-specific context, but practical use is complicated by temporal changes (including pathology progression), misalignment between scans, and protocol drift across acquisitions. In this work, we introduce **L-TGVN**, a **L**ongitudinal **T**rust-**G**uided **V**ariational **N**etwork that leverages prior scans as side information to reconstruct the current scan from heavily undersampled measurements. Crucially, L-TGVN constrains the influence of prior scans to be consistent with the acquired measurements. Unlike many existing longitudinal reconstruction methods, it does not require explicit pre-registration between prior and current scans. It further accommodates differences in acquisition protocols across visits (e.g., changes in sequence parameters). We evaluate L-TGVN against matched-capacity baselines, including prior-guided methods and methods that do not use longitudinal priors, and observe consistent improvements in standard quantitative metrics together with better preservation of fine structures at challenging accelerations. Source code is available at github.com/sodicksonlab/L-TGVN.

**Keywords:** Longitudinal MRI · Temporal context · MRI reconstruction

## 1. Introduction

Magnetic Resonance Imaging (MRI) is a cornerstone of clinical diagnostic imaging, offering ionization-free examinations alongside excellent soft-tissue delineation. An MR scanner acquires measurements in frequency space, called *k*-space, that encode the body's response to applied electromagnetic fields, with multiple receiver coils capturing *k*-space samples that are modulated by coil sensitivities. This process is mathematically described by a linear map called the *forward operator*. The *k*-space measurements are then used to reconstruct a spatially resolved image by solving the corresponding linear inverse problem (LIP). Despite its advantages, MRI is intrinsically slow due to the way spatial encoding is performed. Long scan times increase patient discomfort and susceptibility to motion while also increasing exam cost and limiting scanner throughput. A widely used strategy to accelerate MRI is to acquire fewer *k*-space samples and reconstruct images by solving the corresponding LIP.

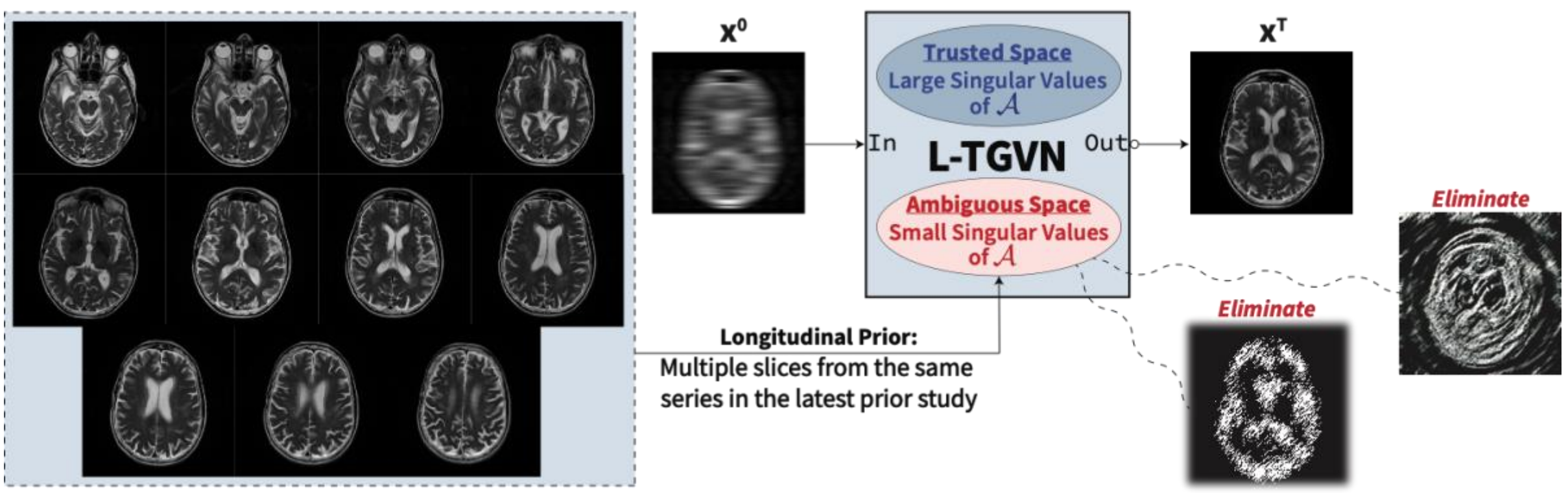

**Fig. 1. Longitudinal trust-guided disambiguation of solutions to the LIP.** As side information, we used 11 axial T2-weighted (T2w) DICOM slices from the most recent prior study to reconstruct a target axial T2w slice in the follow-up exam from its highly undersampled *k*-space measurements.

Beyond certain levels of undersampling, this LIP becomes ill-posed: multiple candidate images can explain the measurements, and small perturbations in measurements can lead to markedly different reconstructions. Recovering diagnostic-quality images then becomes challenging and typically requires additional prior knowledge.

Classical acceleration techniques, including parallel imaging and compressed sensing, encode prior knowledge through coil-sensitivities, sparsity, and transform-domain regularization [11, 13, 18]. More recently, deep learning (DL) has emerged as a dominant paradigm for MRI reconstruction. One common class of DL reconstruction algorithms learns powerful image priors from data while explicitly enforcing the forward model through recurrent unrolled architectures [1, 4, 19]. These approaches can substantially improve reconstruction quality and speed; however, at high accelerations the resulting LIP becomes increasingly ambiguous, leaving more degrees of freedom unresolved and thereby limiting DL methods that rely solely on population priors [14].

A particularly compelling source of additional prior knowledge arises in longitudinal imaging studies. In routine follow-up exams, many patients already have a recent prior scan that provides highly informative, subject-specific structural context (a.k.a. *side information*). If utilized appropriately, such longitudinal priors can disambiguate highly undersampled reconstructions beyond what population-level priors alone can offer. However, practical use is nontrivial: longitudinal scans may differ due to (a) genuine temporal changes, including pathology progression or treatment effects, (b) misalignment due to inter-scan coordinate-frame mismatch (e.g., different slice prescription and patient positioning), and (c) protocol drift or scanner differences that alter intensity statistics and contrast. These factors make naive similarity penalties or direct reference substitution unreliable and, more importantly, raise concerns of diagnostic accuracy: over-reliance on the prior can suppress true interval changes or introduce plausible-looking but incorrect structures, thereby compromising diagnostic reliability.

In this work, we introduce **L**ongitudinal **T**rust-**G**uided **V**ariational **N**etwork (**L-TGVN**) for accelerated MRI with longitudinal side information. L-TGVN incorporates the patient's most recent scan as side information but restricts its influence to remain consistent with the current undersampled *k*-space measurements. This design aims to preserve clinically relevant interval changes while still exploiting the strong subject-specific context available in follow-up studies. Unlike many prior-guided longitudinal approaches that require explicit pre-registration between visits, L-TGVN is designed to operate without a separate registration pipeline and to tolerate protocol differences across acquisitions. We evaluate L-TGVN against matched-capacity baselines, from longitudinally informed approaches to methods that use only the current acquisition, and we observe consistent quantitative improvements together with improved preservation of fine anatomical detail at challenging accelerations.

## 2. Background

*Deep Learning for Multi-coil MR Image Reconstruction*

Let $\widetilde{\boldsymbol{k}}$ denote the fully-sampled *k*-space measurements, which represent Fourier coefficients of the structure of the continuous object being imaged. We define a discrete estimated MR image $\boldsymbol{x}$, such that $\widetilde{\boldsymbol{k}} = \mathcal{F}(\boldsymbol{x}) + \epsilon$, where $\epsilon$ is the measurement noise and $\mathcal{F}$ denotes the Fourier transform operator. In multi-coil acquisition, the scanner captures multiple views of the anatomy modulated by the sensitivities $S_i$ of the receiver coils. Multi-coil *k*-space measurements are given by

$$\widetilde{\boldsymbol{k}}_i = \mathcal{F}(S_i \boldsymbol{x}) + \epsilon_i, \quad i \in \{1,2, \dots, N_c\},$$

where $\epsilon_i$ is the measurement noise for coil $i$ and $N_c$ denotes the number of coils. To simplify notation, we aggregate the *k*-space data from all coils into a single tensor $\widetilde{\boldsymbol{k}} = (\widetilde{\boldsymbol{k}}_1, \dots, \widetilde{\boldsymbol{k}}_{N_c})$ and define the expand operator

$$\mathcal{E}: \boldsymbol{x} \mapsto \big(\mathcal{F}(S_1 \boldsymbol{x}), \dots, \mathcal{F}(S_{N_c} \boldsymbol{x})\big).$$

To accelerate MRI acquisition, fewer *k*-space samples are acquired, which we denote by a binary mask $\mathcal{M}$. The undersampled *k*-space measurements can be denoted as $\boldsymbol{k} \triangleq \mathcal{M}\widetilde{\boldsymbol{k}} = (\mathcal{M}\widetilde{\boldsymbol{k}}_1, \dots, \mathcal{M}\widetilde{\boldsymbol{k}}_{N_c})$, and the forward operator $\mathcal{A}$ is equal to $\mathcal{M} \circ \mathcal{E}$. That is,

$$\boldsymbol{k} = \mathcal{A}\,\boldsymbol{x} + \epsilon' = (\mathcal{M} \circ \mathcal{E})\boldsymbol{x} + \epsilon',$$

where $\epsilon'$ denotes the measurement noise in the undersampled *k*-space. When $\boldsymbol{k}$ is undersampled, the LIP of estimating $\boldsymbol{x}$ from undersampled measurements is often formulated as a regularized least-squares problem, i.e.,

$$\widehat{\boldsymbol{x}} = \arg\min_{x} \frac{1}{2} \|\mathcal{A}\,\boldsymbol{x} - \boldsymbol{k}\|_2^2 + \Psi(\boldsymbol{x}),$$

where $\Psi(\cdot)$ denotes a regularizer that imposes certain constraints on the possible solutions $\boldsymbol{x}$.

*Leveraging Side Information with TGVN*

To leverage patient-specific side information while minimizing the risk of hallucinations, TGVN [2] augments the regularized least-squares objective with a squared Euclidean distance penalty, $\|\mathcal{P}_\delta\, \boldsymbol{x} - \mathcal{H}(\boldsymbol{s}; \gamma)\|_2^2$, where $\mathcal{H}$ is a learnable module parameterized by $\gamma$, and $\mathcal{P}_\delta$ denotes the orthogonal projector onto the subspace spanned by the right singular vectors of $\mathcal{A}$ whose associated singular values $\sigma_i$ satisfy $\sigma_i < \delta$. Starting from the zero-filled reconstruction, TGVN performs $T$ Landweber iterations [8]:

$$\boldsymbol{x}^{t+1} = \boldsymbol{x}^t - \eta^t \mathcal{A}^H(\mathcal{A}\,\boldsymbol{x}^t - \boldsymbol{k}) - \underbrace{\mu^t\,\mathcal{P}_\delta\big(\boldsymbol{x}^t - \mathcal{H}(\boldsymbol{s};\gamma^t)\big)}_{\text{trust guidance}} - \Phi(\boldsymbol{x}^t;\theta^t),$$

where $\Phi(\cdot)$ is a learned regularizer that leverages population priors. By design, $\mathcal{P}_\delta$ confines the influence of side information to the ill-conditioned (ambiguous) subspace, thereby limiting potential inconsistencies with the acquired $k$-space. In other words, the influence of side information is constrained to be data consistent.

### 2.1 Related Work

Leveraging patient-specific priors to accelerate MRI has a long history. Early prior-information-driven approaches formalized how a reference image can be incorporated while controlling its influence to mitigate the generation of phantom structures when the prior is imperfect [20]. In explicitly longitudinal settings, several works studied serial imaging where a baseline scan provides subject-specific context for reconstructing a follow-up scan from undersampled measurements, often by promoting sparsity in the baseline–follow-up difference or using prior-weighted regularization [10, 16]. More broadly, longitudinal priors can be viewed as another form of side information: in principle, many multi-contrast reconstruction frameworks can be adapted to longitudinal imaging by treating a prior scan as the auxiliary input.

A prominent line of work formulated the longitudinal reconstruction through reference-based compressed sensing with mechanisms that adaptively reduce reliance on the prior when longitudinal similarity breaks down [23, 24]. Related optimization-based formulations have also been explored, combining data fidelity with sparsity and explicit penalties on deviation from a reference image [6]. More recently, learning-based longitudinal reconstruction has emerged, including approaches that aim to reduce reliance on paired training data and pipelines that learn to align and fuse prior information [12, 17, 21].

Despite substantial progress, existing methods often face a core tradeoff: tightly prior-coupled approaches can overfit to outdated anatomy or pathology, while looser reliance on priors may underutilize subject-specific information at challenging levels of acceleration, where explicit image-domain registration can also become unreliable. L-TGVN addresses this challenge by constraining the influence of longitudinal priors to remain consistent with the acquired measurements, providing a principled "trust" mechanism that guards against over-reliance while still exploiting subject-specific context. In addition, L-TGVN avoids explicit pre-registration between visits by relying on feature-space fusion within the trust-guidance module. It is designed to accommodate cross-visit acquisition changes as well, making it well suited to practical follow-up imaging workflows.

## 3. L-TGVN: Detailed Description

We consider longitudinal accelerated MRI, where for a given subject the current (follow-up) scan is acquired with undersampling beyond what is achievable by parallel imaging alone. In addition to the current acquisition, we assume access to a subject-specific longitudinal prior $\boldsymbol{s}$, e.g., the most recent scan from a previous visit, as illustrated in Fig. 1. Conceptually, $\boldsymbol{s}$ may include the full prior study, i.e., all image-domain (DICOM) slices across all series. In practice, using the entire prior exam can be infeasible; instead, for each target slice we use a small subset of slices from the prior series that best matches the current acquisition based on metadata. We do not assume that $\boldsymbol{s}$ is pre-registered to the current scan, and we allow $\boldsymbol{s}$ and $\boldsymbol{x}$ to differ due to temporal changes (including pathology progression), misalignment, and protocol drift. Our goal is to reconstruct $\boldsymbol{x}$ from $\boldsymbol{k}$ while leveraging $\boldsymbol{s}$ only to the extent supported by the acquired measurements.

*Longitudinal Trust Guidance*

Unlike the original TGVN formulation where the residual term before the projection can be expressed as $\boldsymbol{r}^t = \boldsymbol{x}^t - \mathcal{H}(\boldsymbol{s}; \gamma^t)$, L-TGVN uses a conditional guidance network that depends on both the longitudinal prior and the current iterate. That is, $\tilde{\boldsymbol{x}}^t = \mathcal{H}(\boldsymbol{s}, \boldsymbol{x}^t; \gamma^t)$ and $\boldsymbol{r}^t = \boldsymbol{x}^t - \tilde{\boldsymbol{x}}^t$.

In our implementation, $\mathcal{H}$ is a dual-encoder, shared-decoder U-Net [15]. One encoder processes the prior input $\boldsymbol{s}$, represented as a small stack of neighboring slices from the most relevant prior series (e.g., $N_s$ slices selected based on metadata), while a second encoder processes the current complex-valued estimate $\boldsymbol{x}^t$, mapped to two real channels. At each resolution, features from the two encoders are fused (concatenation followed by a 1×1 convolution) to form shared skip connections for a single decoder, which outputs a complex-valued proposal $\tilde{\boldsymbol{x}}^t$ with the same spatial support as the current reconstruction. Conditioning on $\boldsymbol{x}^t$ allows $\mathcal{H}$ to learn implicit alignment and protocol adaptation within the network, avoiding an explicit pre-registration pipeline. Starting with the zero-filled, coil-combined reconstruction $\boldsymbol{x}^0 = \mathcal{A}^H \boldsymbol{k}$, we unroll $T$ Landweber iterations of

$$\boldsymbol{x}^{t+1} = \boldsymbol{x}^t - \eta^t \mathcal{A}^H(\mathcal{A}\,\boldsymbol{x}^t - \boldsymbol{k}) - \mu^t\, \mathcal{P}_\delta\, \boldsymbol{r}^t - \Phi(\boldsymbol{x}^t; \theta^t),$$

and obtain $\boldsymbol{x}^T$. Given the target $\boldsymbol{x}^*$, we learn all trainable parameters across iterations in a supervised manner by maximizing a chosen similarity metric (SSIM [22]) between $\boldsymbol{x}^T$ and $\boldsymbol{x}^*$. These parameters include the singular value threshold $\delta$, which the efficient projection implementation renders differentiable [2]. Since $\mathcal{P}_\delta$ confines side information to the ambiguous subspace of $\mathcal{A}$, trust guidance regularizes only poorly constrained components, i.e., unacquired $k$-space regions in single-coil acquisitions, and high $g$-factor directions in multi-coil acquisitions. Furthermore, the coil sensitivities are refined as in [2, 19], guarding against gross inaccuracies. L-TGVN thus inherits these safeguards and should remain non-inferior to a no-side-information baseline under degraded operators.

## 4. Empirical Validation and Results

*Evaluation questions and baselines*

In our experiments, we address two main questions: **Q1)** Is there a benefit to leveraging longitudinal priors, compared to increasing the capacity of the learned regularizer to enforce stronger population priors? **Q2)** How effectively does L-TGVN exploit longitudinal priors? To answer Q1, we compared L-TGVN against E2E-VN [19], which does not use longitudinal priors, and expanded E2E-VN's learned regularizer to match the total parameter count between the two models. To answer Q2, we compared L-TGVN against two matched-capacity, DL-based baselines that leverage longitudinal priors and have publicly available implementations: MTrans [3] and DMSI [9]. Note that these baselines were originally proposed for multi-contrast reconstruction; we adapt them to longitudinal imaging by using the prior scan as the conditioning input. We also performed ablation studies to assess the effect of incorporating multiple series (e.g., T2w and FLAIR) from the prior study. The results favor using multiple series, but we omit the detailed analysis due to space constraints.

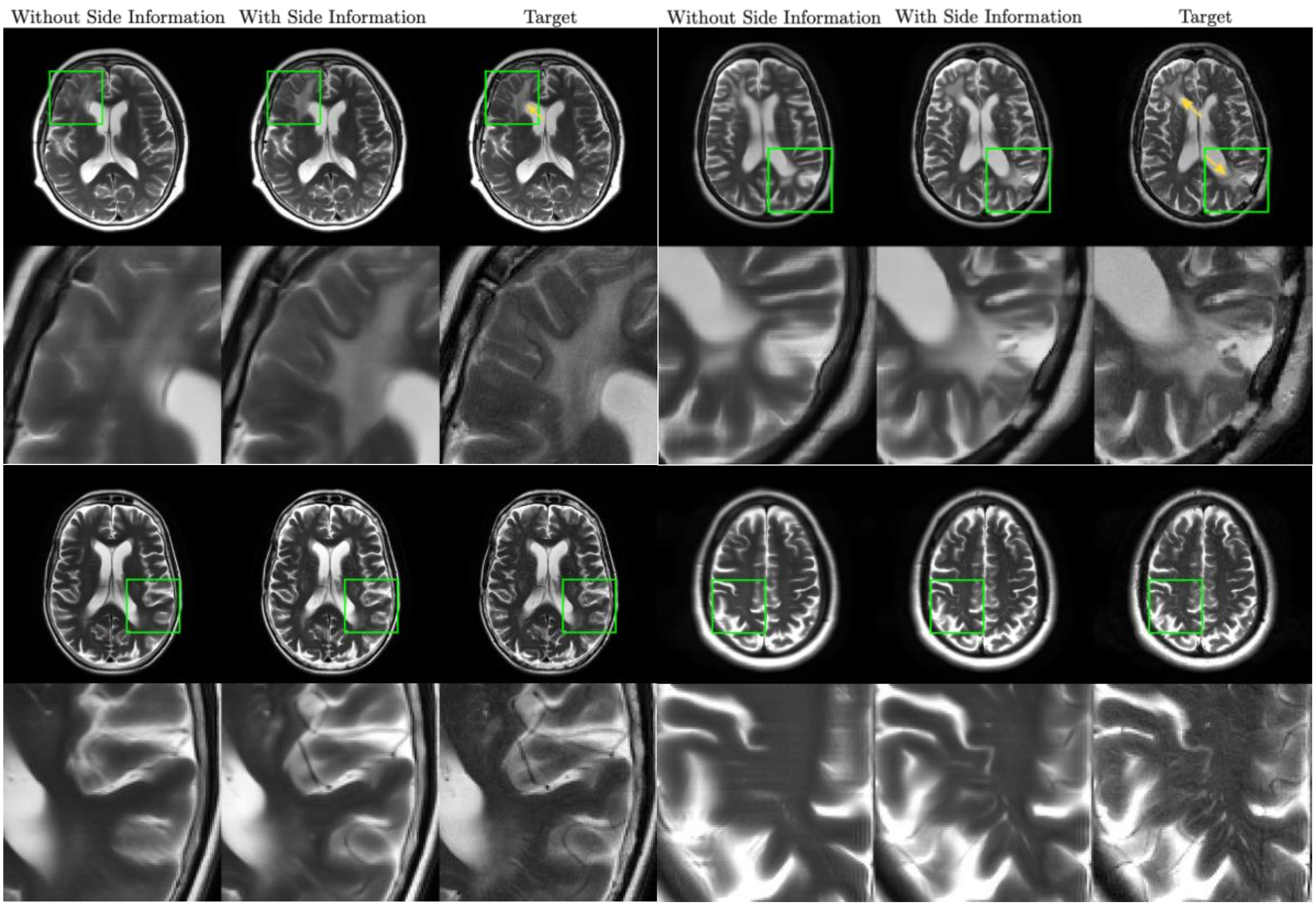

**Fig. 2. Using longitudinal priors significantly enhances reconstruction quality.** Axial T2w reconstructions from the test split (B1, 20× acceleration); green boxes denote the zoomed regions (bottom rows). L-TGVN recovers fine neuroanatomical details and pathologies (yellow arrows), matching the target whereas without it key brain structures are homogenized and the highlighted pathologies become difficult to discern.

*Dataset and longitudinal pairing*

We used an in-house, IRB-approved and de-identified fully sampled multi-coil brain *k*-space dataset comprising 757/91/81 axial T2-weighted (T2w) volumes for training/validation/testing, corresponding to 19,168/2,287/2,087 slices, respectively, with patient-level splits to avoid data leakage. Raw data were acquired on 3T and 1.5T scanners using turbo spin-echo sequences [5]. For each subject, we identified the most recent prior study in our Picture Archiving and Communication System (PACS) and retrieved the axial T2w DICOM series as longitudinal side information, avoiding repeat studies. The median inter-scan interval was 326 days (Q1–Q3: 159–548 days). The dataset is heterogeneous: acquisition parameters, scanner models, and field strengths can differ between the current and prior scans, and we observe appreciable inter-scan misalignment, including in-plane and out-of-plane rotations and translations.

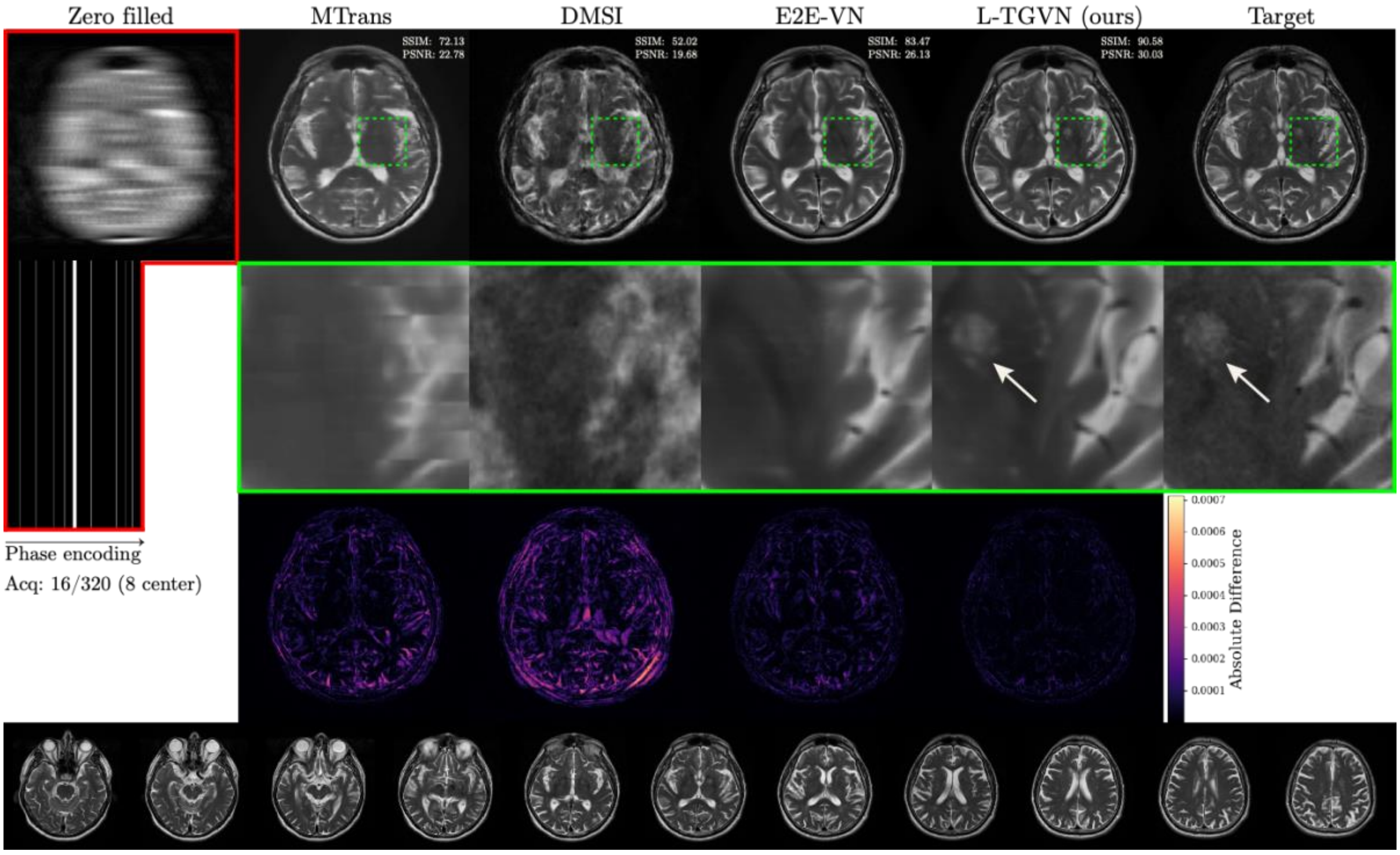

**Fig. 3. Reconstructions from B1 (20×) demonstrating effectiveness of L-TGVN in utilizing longitudinal priors.** The *lesion* (white arrow) is clearly visible only in the L-TGVN reconstruction. **Top:** Full field-of-view images. **Middle:** Undersampling mask, zoomed-in regions, and absolute difference maps. **Bottom:** The 11 slices used as side information, acquired 5 months before the current scan. Note the rigid misalignment between prior and current scans.

*Retrospective undersampling setup*

In experiments B1–B3, we reconstructed axial T2w magnitude images from retrospectively undersampled $k$-space using random phase-encoding (PE) masks with a fully sampled center. We used acceleration factors of 20×, 15×, and 10× for B1, B2, and B3, respectively, with center-frequency fractions of 2.5%, 3.3%, and 5%. In all cases, acceleration was achieved solely by 1D undersampling, and the reported factors denote the net acceleration including the fully sampled center. All longitudinal-prior methods used $N_s = 11$ prior slices, selected based on metadata ($z$-axis position), as side information when reconstructing each target slice from its undersampled $k$-space.

**Table 1. Quantitative evaluation results.** SSIM, PSNR, and NRMSE are reported for the experiments B1–B3 using L-TGVN and the baselines. Values are shown as mean ± standard error over the test set, computed at the patient level ($n = 81$). Bold entries indicate the best performance for each metric.

| Metric | Experiment | MTrans | DMSI | E2E-VN | L-TGVN (ours) |
|---|---|---|---|---|---|
| SSIM | B1 | 77.05 ± 0.49 | 53.78 ± 0.44 | 87.65 ± 0.39 | **90.89 ± 0.36** |
| | B2 | 79.44 ± 0.45 | 54.98 ± 0.42 | 89.46 ± 0.32 | **91.56 ± 0.31** |
| | B3 | 83.94 ± 0.37 | 60.69 ± 0.34 | 92.12 ± 0.28 | **93.05 ± 0.26** |
| PSNR | B1 | 25.02 ± 0.17 | 22.09 ± 0.18 | 29.13 ± 0.22 | **31.27 ± 0.23** |
| | B2 | 26.03 ± 0.16 | 22.72 ± 0.17 | 30.62 ± 0.19 | **32.20 ± 0.21** |
| | B3 | 28.28 ± 0.15 | 24.30 ± 0.14 | 32.84 ± 0.20 | **33.81 ± 0.20** |
| NRMSE | B1 | 0.512 ± 0.007 | 0.717 ± 0.011 | 0.322 ± 0.006 | **0.253 ± 0.006** |
| | B2 | 0.456 ± 0.006 | 0.667 ± 0.010 | 0.270 ± 0.004 | **0.227 ± 0.004** |
| | B3 | 0.351 ± 0.004 | 0.553 ± 0.006 | 0.210 ± 0.004 | **0.188 ± 0.003** |

*Training protocol and model selection*

For a fair comparison, L-TGVN and the baselines were configured to have a similar number of trainable parameters (≈ 500M). L-TGVN, MTrans, and E2E-VN were trained on 8× A100 GPUs with 80 GB VRAM with an equivalent batch size of 64 for 20K gradient steps, with parameters saved at each epoch. The model with the best validation SSIM was selected for testing. DMSI was trained for 70K gradient steps with the same setup as it requires longer training, and the model with the best validation EDM loss [7] was selected for testing. Additional details are provided in the repository.

*Results*

Fig. 2 shows the reconstruction results for axial T2w brain images with (L-TGVN) and without (E2E-VN) using side information. At 20× acceleration, side information significantly aids reconstruction while its absence results in loss of fine anatomical details and pathologies highlighted by the yellow arrows. As shown in Fig. 3, L-TGVN produces the most faithful reconstructions among methods that incorporate longitudinal priors. In contrast, MTrans and DMSI tend to oversmooth the images, washing out fine structures; this failure mode is especially apparent in the absolute difference maps, which highlight widespread residual error. L-TGVN largely avoids this degradation, retaining sharper edges and better preserving subtle anatomical details (e.g., the lesion). Table 1 presents quantitative results, showing that L-TGVN achieves the best performance across all image quality metrics, with significant improvements confirmed by Wilcoxon signed-rank test ($p < 0.01$). Together, Fig. 2 and Table 1 answer Q1 by confirming that longitudinal priors can noticeably enhance reconstruction quality. Meanwhile, Fig. 3 and Table 1 answer Q2 by showing that L-TGVN makes stronger, more reliable use of prior scans than matched-capacity alternatives.

## 5. Conclusion

We introduced **L-TGVN** for accelerated MRI, leveraging a patient's most recent prior scan as subject-specific side information. Our key finding is that longitudinal priors can substantially improve reconstruction quality at challenging accelerations, provided that their influence is constrained to remain data-consistent. This trust-guided design helps preserve fine anatomical detail while reducing the risk of over-reliance on outdated information in the presence of temporal change (e.g., pathology progression), inter-scan misalignment, and protocol drift.

*Limitations and future work*

We focused on using the most recent prior scan as longitudinal side information, represented by a small stack of neighboring slices from the most relevant prior series. In future work, we plan to extend L-TGVN to leverage multiple prior visits and series, conduct radiologist reader studies, and incorporate richer longitudinal context such as radiology reports and clinical metadata. We also plan to explore uncertainty-aware L-TGVN variants to better flag regions where prior information conflicts with current measurements, enabling adaptive acquisition and reconstruction.

**Acknowledgments.** This work was supported in part by the NIH/NIBIB under grant number P41 EB017183, and by the NSF under grant number 1922658.

**Disclosure of Interests.** In December of 2025, not in connection to this work, Dr. Sodickson joined Function Health as Chief Medical Scientist. He also receives royalties from an image reconstruction patent licensed by Siemens Healthineers. The other authors declare no competing interests.